\documentclass[showpacs,preprintnumbers,amsmath,amssymb]{revtex4}

\usepackage{graphicx}% Include figure files
\usepackage{dcolumn}% Align table columns on decimal point
\usepackage{bm}% bold math
\usepackage{color} 
\begin{document}

\title{Quantum Bound States Around Black Holes
}% Force line breaks with \\

\author{J. Grain}
 \email{grain@apc.univ-paris7.fr}
\author{A. Barrau}%
 \email{Aurelien.Barrau@cern.ch}
\affiliation{%
Laboratory for Subatomic Physics and Cosmology,
CNRS-IN2P3 / UJF\\
53, avenue des Martyrs, 38026 Grenoble cedex, France 
}%

%\affiliation{

\date{\today}% It is always \today, today,

\begin{abstract}
Quantum mechanics in the vicinity of black holes is a fascinating
field of theoretical physics. It involves both general relativity and particle
physics, opening new eras to establish the principles of unified theories. In 
this article, we show that quantum bound states with no classical equivalent 
--as can easily be seen at
the dominant monopolar order-- should be formed around black holes for massive 
scalar particles. We qualitatively investigate some important physical 
consequences, in particular for the Hawking evaporation mechanism and the
associated greybody factors. 
\end{abstract}

\maketitle

\section{introduction}

Black holes are extreme objects whose study is very rich and relies on
different branches of physics. Among the most important and fundamental phenomena of black hole
physics are the Hawking evaporation mechanism --see \cite{unruh0} for a review--
and the existence of quasi-normal modes (QNMs) --see \cite{kokko} for a review. On the
one hand, the evaporation phenomenon reveals the profound links between gravity 
and thermodynamics. Its study is both extremely fruitful in itself and because of
the quantum gravitational effects expected to occur during the last
stages of the evaporation, when the semi-classical approach breaks down. On the 
other hand, quasi-normal modes are of
particular importance because they revealed black holes to be stable under
perturbations and represent a key ingredient in the computation of
gravitational wave signals. 

This article focuses on the investigation of bound
states for massive particles emitted by black holes. The problem is
mathematically quite close to the investigation of massive QNMs
\cite{oha}, as in both cases the point is to find the characteristic complex 
frequencies allowing a massive (scalar) field to propagate in a black hole 
background, while satisfying given boundary conditions at the black hole horizon 
and at spatial infinity (see, {\it e.g.}, \cite{dolan}). Nevertheless because
those boundary conditions are not the same, the physical meaning of bound states
and quasi-normal modes are very different and correspond to different energy
ranges. Although bound
states are well known to exist in quite a lot of classical and quantum systems,
this work points out their specific existence around black holes, even
at the monopolar order, and investigates some important consequences. In most of
the
literature, bound states refer both to particles {\it orbiting} the black hole
or back scattered to the black hole. In this work, we are interested in quanta
trapped between the finite black hole potential barrier and the infinitely thick
well which prevents the particles from reaching infinity if their energy is low
enough. As will be demonstrated hereafter, those orbiting states are not 
strictly stable but can be characterized by finite lifetimes. 

In Section II, we first determine the conditions for such states to exist and 
compute their energy spectra and lifetimes at the WKB order. Thanks to a simple 
toy model, we estimate in Section III the average number of trapped particles as
a function of their characteristic energies. Those trapped
massive particles --whose greybody factors must be computed numerically-- will 
also inevitably modify the Hawking radiation spectrum at infinity, which is
computed in this Section. Some conclusions are finally given together with
perspectives.

\section{Orbiting Bound states}
The investigation of orbiting  quantum bound states around Schwarzschild black 
holes requires one to solve
relativistic quantum mechanical equations in a
curved background, while taking into account a non-vanishing mass. To show that 
those states do exist, the Klein-Gordon equation in a Schwarzschild background 
will be shown to exhibit a radial potential containing a local well for given 
ranges of black hole horizon radii and particle masses. 

\subsection{Conditions for such states}
\begin{figure}[ht]
	\begin{center}
		\includegraphics[scale=0.7]{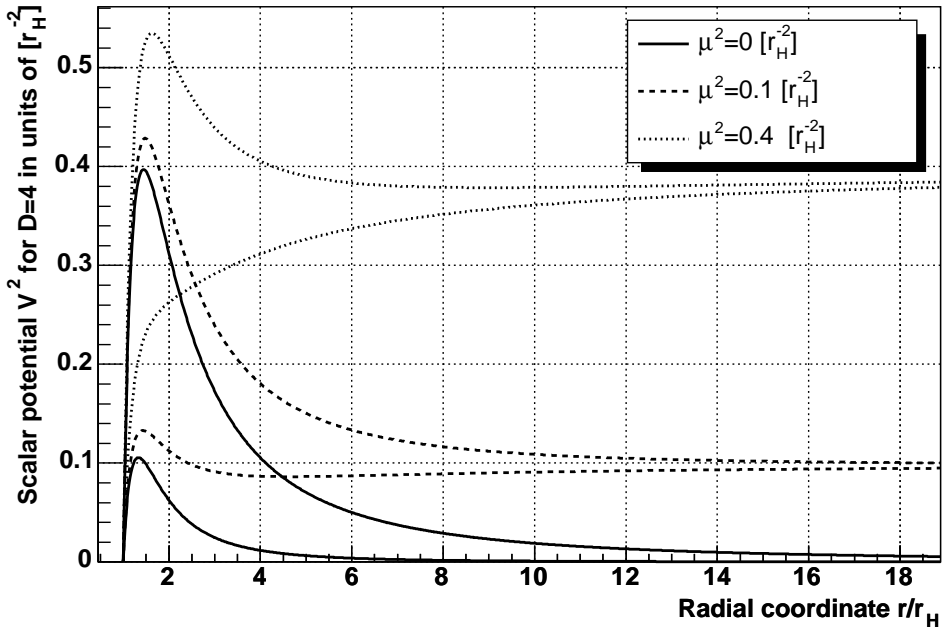}
	\end{center}
	\caption{Square of the potential as a function of the radial coordinate
	for three values of the mass $\mu$ and two values of the angular quantum
	number $\ell$ (in each case, $\ell=0$ for the lower curve and $\ell=1$
	for the upper curve).}
	\label{pot-gen}
\end{figure}

The Klein-Gordon equation of motion for a scalar field $\Phi$ with mass $\mu$ in a
space-time with metric $g_{\alpha\beta}$ can be expressed as
\begin{equation}
\frac{1}{\sqrt{-g}}\partial_\alpha\left(\sqrt{-g}g^{\alpha\beta}\partial_\beta\Phi\right)+\mu^2\Phi=0.
\end{equation}
Writing $\Phi=e^{-i\omega{t}}Y^\ell_m(\theta,\varphi)R(r)$ to split the
temporal, angular and radial parts of the field (where $Y^\ell_m$ are the spherical
harmonics), the radial function $R(r)$ obeys, in a $4-$dimensional Schwarzschild 
background,
\begin{equation}
	\left[\frac{h(r)}{r^2}\frac{d}{dr}h(r)r^2\frac{d}{dr}+\omega^2-h(r)\left(\frac{\ell(\ell+1)}{r^2}+\mu^2\right)\right]R(r)=0,
\end{equation}
where $h(r)$ is defined by the metric $ds^2=h(r)dt^2-dr^2/h(r)-r^2d\Omega^2$ 
(see, {\it
e.g.}, \cite{kanti0} and references therein for a description of the 
general techniques associated with
quantum fields in a Schwarzschild spacetime used throughout this paper).
Under the change of variables $r\rightarrow r_*$ and $R(r)\rightarrow U(r)$ 
where $r_*$ is the tortoise coordinate (such that $dr_*=dr/h(r)$) and
$U(r)=rR(r)$, this equation takes a Schr\"odinger-like form
\begin{equation}
	\frac{d^2U}{dr_\star^2}+\left(\omega^2-V^2_\ell(r)\right)U=0,
	\label{tort}
\end{equation}
with a potential
\begin{equation}
V_\ell^2(r)=\left(1-\frac{r_H}{r}\right)\left(\frac{\ell(\ell+1)}{r^2}+\frac{r_H}{r^3}+\mu^2\right),
\end{equation}
where $r_H$ stands for the Schwarzschild radius and $\ell$ for
the angular quantum number. The usual quantum mechanical
techniques can therefore be employed in the tortoise coordinate system. The 
Chandrasekhar convention is used hereafter : the last term of equation
(\ref{tort}) is interpreted as the squared potential so as to recover the standard
Hamilton-Jacobi equation. On Fig.~\ref{pot-gen},
$V_{\ell}^2(r)$ is shown for three different values of the 
mass ($\mu=\left\{0,\sqrt{0.1},\sqrt{0.4}\right\}~\left[r^{-1}_H\right]$) and 
two values of the angular momentum ($\ell=0$ and $\ell=1$). Depending on $\mu$
and $\ell$, it can be seen that a local minimum, potentially leading to a bound state,
eventually appears. The existence of
a potential well depends on the roots of the algebra\"{\i}c equation 
$\frac{dV^2_\ell}{dr}=0$~:
\begin{equation}
	r_H\mu^2r^3-2\ell(\ell+1)r^2-3r_H(1-\ell(\ell+1))r+4r^2_H=0.
	\label{min-mass}
\end{equation}
Two roots above $r_H$ exist if the mass $\mu$ is lower than a critical value
$\mu_+(\ell)$ given by
\begin{equation}
	\mu^2_+=\frac{1}{216r^2_H}\left(-27J_1+\sqrt{729J^2_1+432J_2}\right),
\end{equation}with
\begin{eqnarray}
	J_1&=&\ell^3(\ell+1)^3+\ell^2(\ell+1)^2-\ell(\ell+1)-1, \\
	J_2&=&\ell^2(\ell+1)^2\left[9+14\ell(\ell+1)+9\ell^2(\ell+1)^2\right].
\end{eqnarray}
In the monopolar case ({\it i.e.} $\ell=0$), it takes the simple value $\mu_+=\frac{1}{2}r_H^{-1}$. 
This allows one to understand easily the general behavior of the potential. 
Let us consider a particle with a given mass $\mu_1$ and a black hole with a 
horizon radius $r_1$. The mass of the particle may be smaller than $1/2r_1$. 
In this case, the local barrier and the potential well do exist for all the 
values of the orbital quantum number, and particles can be potentially trapped 
with any angular momentum. On the other hand, if $\mu_1$ is greater than 
$1/2r_1$, then the $\ell$-domain has to be divided into two subclasses. 
A critical value of the orbital quantum number, denoted $\ell_1$, can then be 
defined by $\mu_+(\ell_1)<\mu<\mu_+(\ell_1+1)$. For all the partial waves with
 an orbital quantum number smaller or equal to this critical value, the 
 potential appears as a monotonically increasing function of $r$ from $V_\ell(r_1)=0$ to $V_\ell(+\infty)=\mu_1$. However, all the partial waves with $\ell>\ell_1$ will face the local barrier as well as the well potential.

%For example, if a
%particle has a mass $\mu\in [\mu_+(\ell_1),\mu_+(\ell_1+1)]$, then the potential is an
%increasing bijection from $[r_H,\infty[$ to $[0,\mu[$ for every multipolar order
%lower or equal to $\ell_1$ whereas a potential barrier and a local well appear
%for multipolar orders strictly higher than $\ell_1$.

The detailed shape of the potential is also determined by another critical mass,
hereafter called $\mu_-(\ell)$, which defines the relative height of the potential
barrier close to the horizon when compared with the mass of the particle. If the
mass of the particle is higher than the barrier, there is no more a turning point 
for the system. The mass is the highest value of the potential if the equation
$V^2_\ell(r)=\mu^2$ has no root above $r_H$. The cubic form
\begin{equation}
	r^3-\frac{\ell(\ell+1)}{r_H\mu^2}r^2-\frac{1-\ell(\ell+1)}{\mu^2}r+\frac{r_H}{\mu^2}=0
	\label{turn-mass}
\end{equation}
satisfies this criterion if $\mu>\mu_-(\ell)$, which is then given by
\begin{equation}
	\mu^2_-=\frac{1}{27r^2_H}\left(-L_1+\sqrt{L^2_1+27L_2}\right),
\end{equation}
with
\begin{eqnarray}
	L_1&=&2\ell^3(\ell+1)^3+3\ell^2(\ell+1)^2-3\ell(\ell+1)-2, \\
	L_2&=&\ell^2(\ell+1)^2\left[\ell(\ell+1)+1\right]^2.
\end{eqnarray}
In the monopolar case, it becomes $\mu_-=\frac{2}{\sqrt{27}}r_H^{-1}$. Table \ref{mass} provides the values of the critical masses in units of 
$r^{-1}_H$.

\begin{table}[ht]
	\begin{center}
		\begin{tabular}{l||c|c}
		$\ell$ & $\mu_-~\left[r^{-1}_H\right]$ & $\mu_+~\left[r^{-1}_H\right]$ \\ \hline\hline
		0 & $\frac{2}{\sqrt{27}}$ & $\frac{1}{2}$ \\
		1 & 0.794 & 0.931 \\
		2 & 1.275 & 1.480 \\
		3 & 1.768 & 2.046 \\
		4 & 2.264 & 2.617 \\
		5 & 2.761 & 3.191
		\end{tabular}
	\end{center}
	\caption{Critical masses in units of $r^{-1}_H$ for different angular
	quantum numbers.}
	\label{mass}
\end{table}

\begin{figure}[ht]
	\begin{center}
		\includegraphics[scale=0.7]{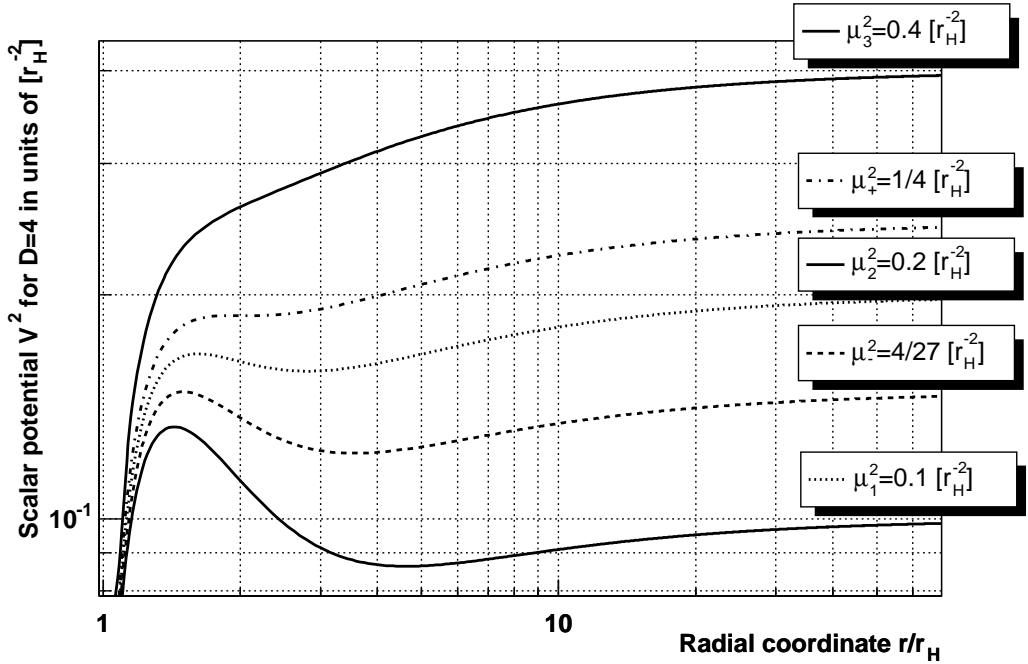}
	\end{center}
	\caption{Squared monopolar potential for five particle masses~: i) $\mu^2=0.1~r^{-2}_H$ 
	corresponding to $\mu<\mu_-(0)$, ii) $\mu=\mu_-(0)$, iii) $\mu^2=0.2~r^{-2}_H$ 
	corresponding to $\mu_-(0)<\mu<\mu_+(0)$, iv) $\mu=\mu_+(0)$ and v) $\mu^2=0.4~r^{-2}_H$
	corresponding to $\mu>\mu_+(0)$.}
	\label{pot-crit}
\end{figure}

The potential is always  zero at $r=r_H$ and tends to $\mu$ for $r\rightarrow
\infty$. i) If $\mu<\mu_-(\ell)$, the potential reaches a maximum higher than
$\mu$
and then reaches a minimum; if ii) $\mu=\mu_-(\ell)$, the maximum is exactly equal
to $\mu$ and a minimum also appears; if iii) $\mu_-(\ell)<\mu<\mu_+(\ell)$ the potential
reaches a maximum lower than $\mu$ and still admits a minimum, whereas if iv)
$\mu\geq \mu_+(\ell)$ the potential is a monotonically increasing function of $r$.
This behavior is illustrated for $\ell=0$ on Fig. \ref{pot-crit} for masses corresponding to
those four specific cases. As a direct consequence of the wave description of particles in quantum mechanics, it can be seen that bound states --due to the local
minimum-- can appear at the monopolar order, with {\it no classical equivalent}.
Although particles can of course be classically trapped around a black hole, 
no state without angular momentum  can be found if the quantum behavior is not 
taken into account. Just as in classical mechanics, it is the non-vanishing mass that leads to the trapping (even at the quantum level), but with specific quantum features, lying in the quantization of the angular momentum, and furthermore allowing for trapping even without any angular momentum.
Furthermore, whatever the mass of the particle, a bound state will appear for
high enough multipolar orders so that $\mu<\mu_+(\ell)$. This makes this phenomenon
of "particle trapping" quite generic. Those bound states are described by 
quasi-stationary quantum states that cannot reach spatial infinity but can still
make a transition to the black hole by tunneling through the gravitational barrier.

\begin{figure}[ht]
	\begin{center}
		\includegraphics[scale=0.7]{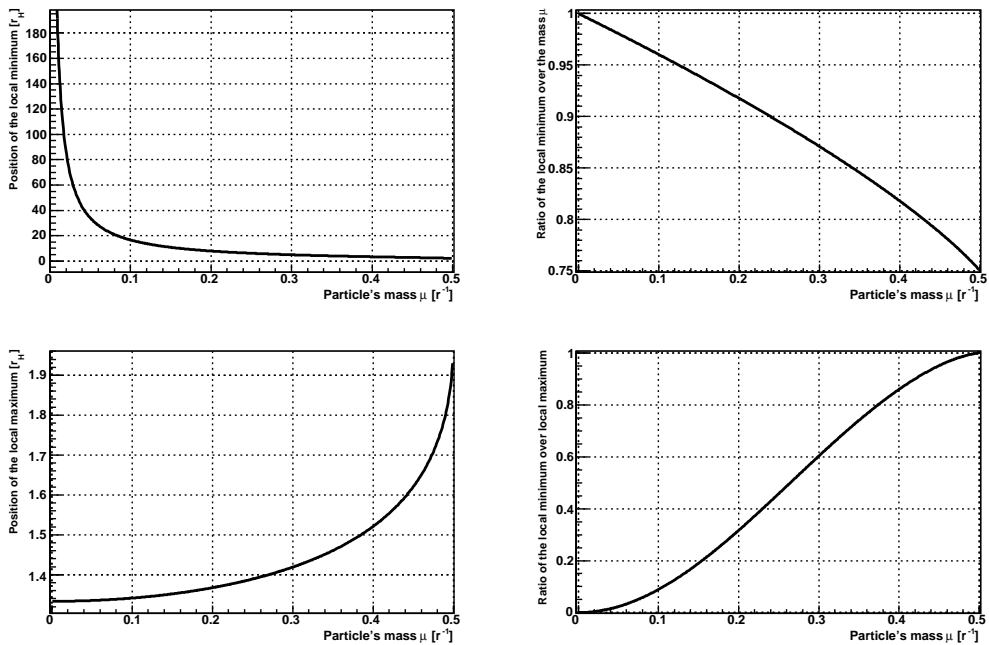}
	\end{center}
	\caption{{\it Upper left~:} position of the local minimum of the potential $V_{\mathrm{min}}$, in units of $r_H$, as a function of the mass of the particle. {\it Upper right~:} ratio of the local minimum over the mass of the particle, $V_{\mathrm{min}}/\mu$, as a function of the mass of the particle. {\it Lower left~:} position of the local maximum of the potential $V_{\mathrm{max}}$, in units of $r_H$, as a function of the mass of the particle. {\it Upper right~:} ratio of the local minimum over the local maximum, $V_{\mathrm{min}}/V_{\mathrm{max}}$, as a function of the mass of the particle.}
	\label{pot-rat}
\end{figure}

As the qualitative features can easily be understood at the monopolar order, this
particular value of the quantum angular momentum is now assumed. When $\mu<\mu_+$, the positions of the potential
barrier ($r_-$) and of the local minimum ($r_+$) can be analytically determined
to be
\begin{equation}
	r_\pm=\frac{1}{\mu}\left[\cos{\left(\theta\right)}\pm\sqrt{3}\sin{\left(\theta\right)}\right], 
\end{equation}
with
\begin{equation}
	\theta=\frac{1}{3}\mathrm{arctan}\left[\sqrt{\left(\frac{\mu_+}{\mu}\right)^2-1}\right].
\end{equation}
The asymptotic behavior is in agreement with the monopolar potential for a massless
particle~:
\begin{equation}
\lim_{\mu\to0}{r_+}=+\infty~,~\lim_{\mu\to0}{r_-}=\frac{4}{3}r_H~,~\lim_{\mu\to
\mu_+}{r_\pm}=2r_H.
\end{equation}
The latter case, $\mu\to \mu_+$, corresponds to a saddle point
at
$r=2r_H$, which represents the degeneracy of the maximum and minimum of the
potential. The most important properties of the monopolar potential are summarized 
in the various graphs of Figure \ref{pot-rat}. As long as the particle's mass is smaller than $\mu_{+}$, 
the potential displays a local well and this feature may lead to the existence of quasi-stationary 
states localized in the well.

\subsection{Complex frequency spectrum : a WKB analysis}
\begin{figure}[ht]
	\begin{center}
		\includegraphics[scale=0.7]{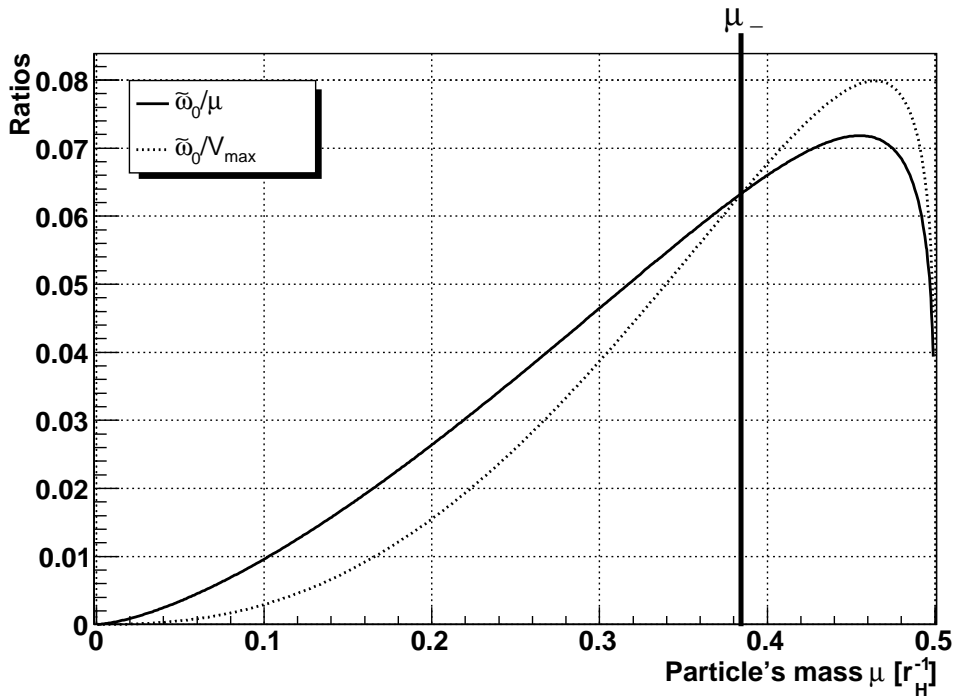}
	\end{center}
	\caption{Ratio of the zero-point energy $\tilde{\omega}_0$ and 
	the mass $\mu$ and ratio of $\tilde{\omega}_0$ and the maximum of the 
	potential $V_{max}$, both as a function of
	the mass of the particle in units of $r_H^{-1}$.}
	\label{zero-quadr}
\end{figure}

Even when there is a well in the potential, this does not guarantee orbiting bound 
states to exist. To ensure the stability of those states, the local well has to be deep enough,
because of the zero-point energy associated with quantum systems. Under the reasonable assumption 
that the shape of the potential
is well approximated by a second order expression, the
dynamics is similar to an harmonic oscillator with a frequency $\tilde{\omega}$ given by the curvature 
of the potential around its minimum :
$
	\tilde{\omega}=\sqrt{\left.d^2V^2/d{r_\star}^2\right|_{r_+}}.
$
This curvature must be evaluated as a function of the tortoise 
coordinate, because this is the coordinate system in which the radial part of the
Klein-Gordon equation is of the Schr\"odinger type. The zero-point energy is simply
given by $\tilde{\omega}_0=\frac{\tilde{\omega}}{2}$. The ratios of this 
approximated computation to the mass of the particle and to the maximum of the 
potential are plotted on Fig. \ref{zero-quadr}. Each time a minimum does appear, the
zero-point energy remains smaller than the mass and the gravitational potential
barrier, allowing bound states to exist. 

Because the left part of the potential well is not an infinite barrier, the orbiting 
bound states are describes by quasi-stationary states associated with resonances in the density of states. 
Those resonances are characterized by their complex frequencies, the real part corresponding to the energy position 
and the imaginary part to the bandwidth. The spectrum of complex frequencies can be determined at the WKB 
order using the techniques developed
in \cite{wkb}. The spectrum will be infinite if $\mu<\mu_-$ and finite if
$\mu_-<\mu<\mu_+$. The Bohr-Sommerfeld rule,  whose validity in a
relativistic framework was established in \cite{wkb}, reads
$\forall~n\in\mathbb{N},~\displaystyle\int^{r_2(\omega)}_{r_1(\omega)}\sqrt{\omega^2-V^2(r)}\frac{dr}{h(r)}=\left(n+\frac{1}{2}\right)\pi$,
the highest frequency allowed for a bound state being $\mu$ for $\mu<\mu_-$ and $V(r_-)$
for $\mu>\mu_-$. In the latter case, the left-hand side integral is clearly finished and
there exist $n_{max}$ states. If $\mu<\mu_-$, the upper bound of the integral is infinite
when $\omega=\mu$ and, the function to be integrated being proportional to $r^{-1/2}$ near
$+\infty$, the spectrum is expected to be infinite. With the appropriate change of
variables in the cubic equation giving the turning points, it can be shown that the
normal frequency spectrum of the resonances depends only on $\mu r_H$ and can be written
as
$\left\{\omega_n\right\}_{n\in\mathbb{N}}=\left\{\frac{f_n(\mu{r}_H)}{r_H}\right\}_{n\in\mathbb{N}}.$
Table \ref{spect-mass} gives some normal frequencies and the associated
bandwidths, as
numerically obtained at the semi-classical order, following \cite{iyer} to evaluate the
tunnel probability. The quantity $N_{eq}$ in this table will be explained in the next section.
\begin{table}
\begin{center}
		\begin{tabular}{c||c|c|c|c|c} \hline\hline
		& \multicolumn{5}{c}{$\mu=0.25\times\mu_+$} \\ \hline\hline
		$n$ & 0 & 1 & 2 & 3 & 4  \\ \hline
		$\omega{r}_H$ & 1.24$.10^{-1}$ & 1.25$.10^{-1}$ & 1.25$.10^{-1}$& 1.25$.10^{-1}$ & 1.25$.10^{-1}$ \\ \hline
		$\Gamma{r}_H$ & 1.49$.10^{-2}$ & 1.49$.10^{-2}$ & 1.50$.10^{-2}$& 1.50$.10^{-2}$ & 1.50$.10^{-2}$  \\ \hline
		$N_{eq}r^{-1}_H$ & 2.15 & 2.09 & 2.09 & 2.09 & 2.09 \\ \hline\hline
		\multicolumn{6}{c}{} \\ \hline\hline
		& \multicolumn{5}{c}{$\mu=\mu_-$} \\ \hline\hline
		$n$ & 0 & 1 & 2 & 3 & 4  \\ \hline
		$\omega{r}_H$ & 3.75$.10^{-1}$ & 3.83$.10^{-1}$ & 3.84$.10^{-1}$ & 3.84$.10^{-1}$ &3.85$.10^{-1}$   \\ \hline
		$\Gamma{r}_H$ & 1.68$.10^{-1}$ & 1.87$.10^{-1}$ & 1.90$.10^{-1}$ & 1.91$.10^{-1}$ & 1.92$.10^{-1}$  \\ \hline
		$N_{eq}r^{-1}_H$ & 0.024 & 0.021 & 0.021 & 0.021 & 0.020 \\ \hline\hline
		\multicolumn{6}{c}{} \\ \hline\hline
		& \multicolumn{5}{c}{$\mu=0.8\times\mu_+$} \\ \hline\hline
		$n$ & 0 & 1 & 2 & 3 & 4  \\ \hline
		$\omega{r}_H$ & 3.88$.10^{-1}$ & / & / & / & /  \\ \hline
		$\Gamma{r}_H$ & 1.91$.10^{-1}$ & / & / & / & /  \\ \hline
		$N_{eq}r^{-1}_H$ & 0.019 & / & / & / & / \\ \hline\hline
		\end{tabular}
	\end{center}
	\caption{First and second rows below the integer $n$ : Spectrum of normal frequencies $\omega$ and 
	bandwidths
	$\Gamma$ evaluated at the WKB order
	in units of $r^{-1}_H$. Third row : Number of trapped particles $N_{eq}$ in the different energy 
	levels once the equilibrium regime is reached (see the text). Three different masses are considered : $\mu<\mu_-$, $\mu=\mu_-$ et $\mu>\mu_-$. }
\label{spect-mass}
\end{table}

\subsection{Qualitative features of the bound states and comparison with quasi-normal modes}
Before investigating the consequences of bound states for Hawking radiation, some 
qualitative features of those states, related to QNMs, and the existence of a halo 
of trapped particles around black holes, are briefly discussed in this section.

As underlined in \cite{dolan}, the computations of bound 
states and QNMs are very similar. The latter are given by the {\it pure outgoing} boundary conditions 
\cite{simone} while the former require evanescent waves at spatial infinity and ingoing modes at the 
black hole event horizon. However, there are no other fundamental physical links between QNMs and bound states. 
First of all, quasi-normal modes are related with the energy carried out by perturbations of black holes 
whereas bound states correspond to quantum states that cannot escape from the gravitational attraction. 
In addition\cite{simone}, the mass $\mu$ of the field has to be smaller than
$\mu_{-}(\ell)$ for QNMs to exist at the $\ell$th multipolar order. Moreover, because quasi-normal 
modes correspond to resonances near the peak of the potential, their energies have to be greater than the mass of the field. 
Those characteristics are direct consequences of the {\it pure outgoing} boundary conditions. The situation is fundamentally
different for orbiting bound states. The condition on $\mu$ is less restrictive : $\mu$ has to be smaller than 
$\mu_{+}(\ell)>\mu_{-}(\ell)$ for orbiting bound states to exist at the $\ell$th multipolar order. Furthermore, 
the energy of
those states is lower than the mass of the particle, due to the gravitational binding energy.

In addition, as long as $\mu<\mu{+}(0)$, bound states exist for $\ell=0$ and a spherical halo of
quanta "orbiting" the black hole can be expected. If the mass is
between $\mu_+(\ell-1)$ and $\mu_+(\ell)$, bound states will exhibit an angular
distribution dominated by the lowest multipolar order allowing for a minimum in
the potential, that is with a distribution roughly given by
$Y^{\ell}_m(\theta,\varphi)$. Clearly, the mass of the field has to be close to the critical masses $\mu_\pm$ for the orbiting 
bound states to have a substantial influence. In particular, if $\mu\ll\mu_+(0)$, the local well is so tiny at any multipolar 
order that particles can be treated as massless. So as to fix the orders of magnitude, table
\ref{part-mass} gives the masses and temperatures black holes should have so
that the effects studied in this article become important for some
{\it standard model} particles. Although they are not spinless --therefore requiring
one to investigate the master equation for fermions and gauge bosons-- the
main qualitative features can safely be inferred from the scalar case.
\begin{table}[ht]
	 \begin{center}
		\begin{tabular}{l||c|c||c|c}
			particle& \multicolumn{2}{c||}{$\mu=\mu_-$ for $\ell=0$} &
			\multicolumn{2}{c||}{$\mu=\mu_+$ for $\ell=0$} \\
			 (mass) &  $M_{BH}$ [kg] & $T_H$ [MeV]  & $M_{BH}$ [kg] & $T_H$ [MeV] \\ \hline\hline
			electron & $9\times10^{13}$ & 0.1 &  $1.2\times10^{14}$ & 0.08 \\
			($511$ keV)  & & & & \\ \hline
			muon & $5\times10^{11}$ & 21.7 &  $6\times10^{11}$ & 16.7 \\
			($105$ MeV)  & & & & \\ \hline
			charm & $4\times10^{10}$ & 248 &  $5\times10^{10}$ & 190 \\
			($\simeq1.3$ GeV)  & & & & \\ \hline
			top &  $2.5\times10^{8}$ & $36\times10^3$  & $3.7\times10^{8}$ & $28\times10^3$ \\
			($171$ GeV)  & & & & \\ \hline
			W boson & $6\times10^{8}$ & $16\times10^3$  & $8\times10^{8}$ & $12\times10^3$ \\
			($80$ GeV)  & & & & \\ \hline\hline
		\end{tabular}
	\end{center}
	\caption{Masses and temperatures expected for a black hole to have the
	critical masses $\mu_{\pm}$ close to the masses of some {\it standard
	model}
	particles.}
	\label{part-mass}
\end{table}

Most {\it very} massive particles (like the top quark or W boson) are
unstable and will decay before reaching infinity or creating a substantial halo
around the hole. This point should, however, not prevent us from considering
bound states associated with lighter stable particles, as the relevant parameter
for the bound states to exist is not the  mass of the field in itself, but the hierarchy
between this mass and the mass of the hole. As far as the field is not strictly
massless, there will be a black hole masse range where $\mu$ becomes close to
$\mu_+$ and $\mu_-$, making the trapping effective.

An important issue to address is related to the time stability of those bound
states. By integrating the Hawking instantaneous energy spectrum and
summing over all the degrees of freedom of the standard model of particle
physics, it is easy to show that the mean time between the emission of two
quanta is, in Planck units, $\Delta t \approx 100 M$. If we consider, {\it
e.g.}, electrons emitted by a black hole such that $\mu \approx \mu_-$, this
leads to $\Delta t \approx 10^{-20}$s; this remains tiny when compared to the
time-scale of the black hole evolution. In other words, as the
energy carried out from the hole by each emitted particle $(E\approx T =
1/(8\pi M))$ is much smaller than the mass of the hole, many quanta can be
emitted and trapped before the structure of the potential will have
substantially changed. This is, of course, not true anymore when $T \sim M$, that
is to say in the Planck regime in which the semi-classical approximation breaks down
anyway.

\section{Consequences for the Hawking radiation}

\subsection{Trapped particles}
The existence of bound states will play an important role in the Hawking evaporation
mechanism which was initially described in \cite{Hawking} without taking into account
this phenomenon.

First, some low-energy particles will be trapped by the potential well. They will not reach
infinity and the spectrum will be modified in a way quite similar to what could happen
due to a QCD halo \cite{heckler}. When the black hole evaporates, new bound states will
appear each time the temperature becomes of the same order than the critical masses
$\mu_{\pm}$ associated with existing particles. At the first level of approximation, the number of particles within the well
per unit of time and energy, $dN/dt$, at a given energy is given by $F_{in}-F_{out}$ with
$F_{in}=\omega/[\tau(e^{\omega/T}\pm1)]$ representing the ingoing one due to the Hawking evaporation, and 
$F_{out}=N/\tau$ the outgoing one. The energy-dependent time constant $\tau=1/\Gamma$ simply corresponds to the lifetime 
of the bound state. This simple model for the number of particles trapped in a local well is realistic only 
in the case of heavy black holes which can be considered as stationary (the small amount of energy carried out by emitted
particles being much smaller that the mass of the black hole). With $N_0$ the initial amount of trapped particles, 
the time evolution is simply given by
\begin{equation}
	N(\omega,t)=N_0(\omega)e^{-t/\tau}+\frac{1}{\omega\left(e^{-\omega/T_H}-1\right)}\left(1-e^{-t/\tau}\right).
\end{equation}
which can be clearly understood as competition between particles "leaking" from the well to the hole and particles
"filling" the well because of the Hawking radiation. Quickly the number of trapped particles 
reaches an equilibrium regime with $N_{eq}\simeq\frac{1}{\omega\left(e^{-\omega/T_H}-1\right)}$. It is 
worth noticing that once the equilibrium regime is reached, the number of trapped particles 
does not depend on $\tau$ but only on the energy and on the temperature of the black hole. 
This behavior is due to the fact that the probability to cross the potential barrier is the same for 
incoming and outgoing particles. As a consequence, the equilibrium regime, given by the ratio of the number 
of particles scattering the off barrier from the left and the number of particles scattering off the barrier from the right, does
not depend on the lifetime of the bound states.

It is important to compare the characteristic time to reach this regime with the characteristic evolution
time of the black hole. For monopolar bound states, which are 
dominant, and for particles with masses below $\mu_-$, the lifetime ranges between $10M$ and $100M$. 
This time-scale is of the same order as the typical time-scale between two successive 
emissions of a particle ,which, as already mentioned in this article, is much
smaller than the evolution time-scale for
the black hole itself. This makes meaningful our hypothesis of {\it non-evolutionary black holes} : 
as long as heavy black holes are considered, enough particles are emitted to 
reach the equilibrium regime without any substantial modifications of the properties of the black hole. The expected
number of trapped particles when the equilibrium regime is reached is also 
given in Table \ref{spect-mass} in units of $r^{-1}_H$, per unit of energy and time. As can be seen from this 
table, 
the mean number of particles decreases as the mass of the particle increases,
since higher energy bound states are involved.

\subsection{Greybody factors and radiation spectra}
\begin{figure}[ht]
	\begin{center}
		\includegraphics[scale=0.7]{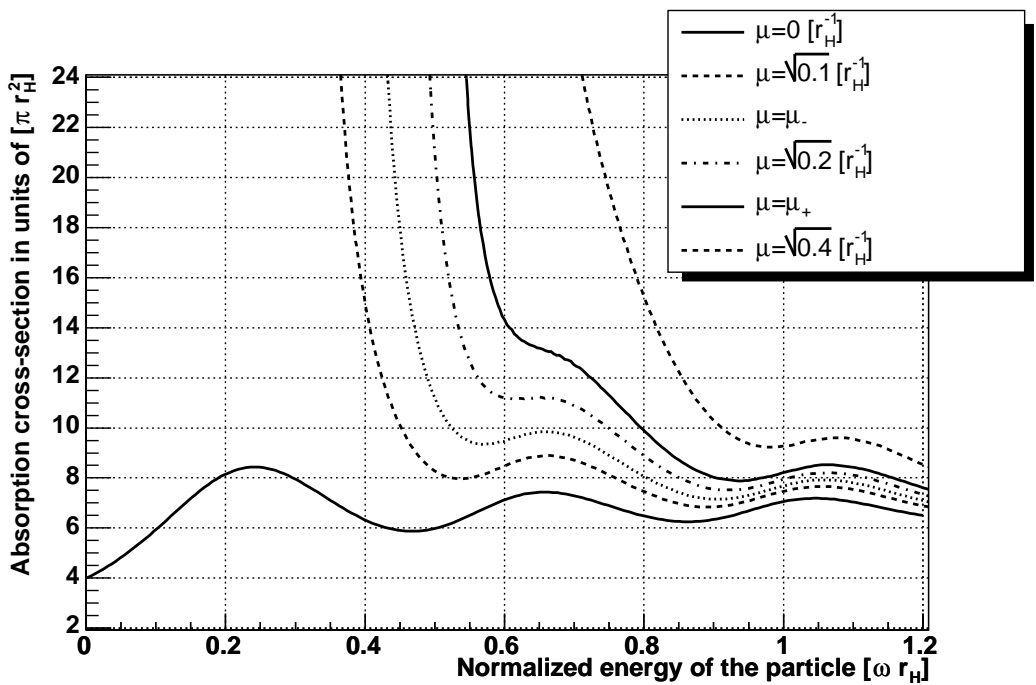}
	\end{center}
	\caption{Absorption cross section for massive scalar particles in units of $\pi
	r_H^2$ as a function of the energy measured at infinity.}
	\label{grey-num}
\end{figure}

The mass of the particle will also drastically modify the greybody factors that
account for the non-trivial part (gravitational barrier and centrifugal potential) of the
couplings between quantum fields and evaporating black holes. The greybody factors (whose
detailed study began with \cite{staro1,staro2} followed by \cite{page1,page2,page3}) have
recently been computed in quite a lot of interesting situations~: extradimensions
\cite{kanti1}, de-Sitter spacetime \cite{kanti2}, rotating black holes 
\cite{kanti3,kanti4}, Gauss-Bonnet gravity \cite{barrau,kanti5} etc., but up to now the
masses of the emitted particles have mostly been ignored (although some good estimates were
obtained in \cite{unruh,page3}). 
Figure \ref{grey-num} displays the absorption cross section
numerically computed by solving the Klein-Gordon equation to evaluate the ingoing and
outgoing amplitudes of the wave function at the horizon and at spatial infinity (see, {\it
e.g.}, \cite{kanti2} for a detailed description of the method we have developed). It should be pointed out that those cross sections have been computed for particles with an energy greater than or equal to the mass. Particles with an energy smaller than the mass will face an infinitely thick potential barrier, precisely due to the mass term in the potential, and will never reach spatial infinity. It should also be noticed that,
when the wavelength of the particle becomes infinite, the cross section
diverges, potentially leading to an experimentally relevant enhancement of soft quanta. It can
indeed be expressed as
\begin{displaymath}
	\sigma_g(\omega)=\displaystyle\sum_\ell\frac{\pi(2\ell+1)}{k^2}\left|A_\ell\right|^2,
\end{displaymath}
where $k$ is the momentum (so that $\omega^2=k^2+\mu^2$ at spatial infinity) 
and $\left|A_\ell\right|^2$ is
the transmission coefficient. However, the transmission coefficient remains 
always smaller than or equal to unity and the divergence lies in {\it geometrical 
issues}. When $\omega\to\mu$, the wave number of the particle tends to zero, 
whereas the transmission coefficient is non-vanishing, just because the 
potential barrier is not infinitely thick (the barrier precisely stops where
the potential well starts and the potential tends asymptotically to $\mu$ with negative values). This leads to a $k^{-2}$ divergence. The situation is clearly different for massless particles. In this case, the potential barrier tends asymptotically to zero with {\it positive values}, leading to $\left|A_\ell\right|^2\propto{k}^2$ when $\omega\to0$. 
This feature prevents the IR divergence from occurring for massless particles. Those results are in agreement 
with analytical investigations in the IR regime,
$\sigma_g(k\to0)\sim4\pi^2(\mu{r}_H)^3/k^2$, as obtained following \cite{unruh}.

\begin{figure}[ht]
	\begin{center}
		\includegraphics[scale=0.7]{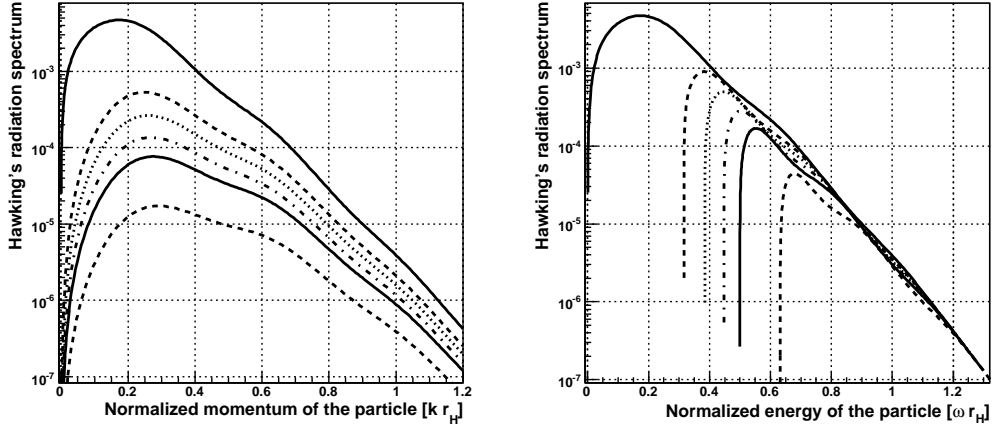}
	\end{center}
	\caption{Flux at infinity emitted by a black hole when taking into account the mass of the
	emitted particle. The different curves share the
	convention of Fig.~\ref{flux}~:
	$\mu=0~[r_H^{-1}],\sqrt{0.1}~[r_H^{-1}],\mu_-,\sqrt{0.2}~[r_H^{-1}],\mu_+~[r_H^{-1}],\sqrt{0.4}~[r_H^{-1}]$ from top
	to bottom. {\it Left panel~:} Hawking's radiation spectrum, $d^2N/dtdk$, as a function of the particle's momentum $k$. {\it Right panel~:} Hawking's radiation spectrum, $d^2N/dtd\omega$, as a function of the energy $\omega$ of the particle.}
	\label{flux}
\end{figure}

	In Fig~\ref{flux}, the flux at infinity
emitted by a black hole is plotted when the masses of the emitted quanta are taken into
account. As can be seen, this substantially modifies the usual picture both because of
the intrinsic cutoff imposed by the mass and because of more subtle effects included in
this analysis, like the selection induced on the allowed quantum multipolar orders 
of the outgoing particle. From Fig~\ref{flux}, it can be seen that the radiation flux decreases for higher masses. However, in the IR limit, this tendency should change, as can be seen using
the analytical limit 
\begin{displaymath}
	\frac{d^2N}{dtdk}(k\to0)\sim\frac{4\pi^2(\mu{r}_H)^3}{e^{\mu/T}-1}
\end{displaymath}
obtained from \cite{unruh}. The flux is non-vanishing only if the
mass of the particle is non-vanishing, enhancing the emission of ultra-soft 
quanta, just as for Schwarzschild-de Sitter black holes \cite{kanti2}.

\section{conclusion}

This study establishes the existence of new bound states around black holes which, at
least at the dominant monopolar order, have no classical equivalent. Although
such states are known to exist in other physical systems, this opens new
perspectives to investigate the detailed features of the Hawking spectrum (with
possible cosmological consequences related, {\it e.g.}, to the primordial power spectrum 
--see \cite{barrau3} for recent limits and \cite{carr} for a review), the intricate
shape of the greybody factors and the propagation of massive quantum fields in the
vicinity of a black hole. Not only could the phenomenology be revised as the
spectra should be quantitatively modified but fruitful thought experiments
associated with light black holes should also take into account those states. 
The intricate problem of backreaction should however be addressed and requires an
exhaustive study in itself.

\end{document}